\documentclass[prb,twocolumn,superscriptaddress]{revtex4}
\usepackage{natbib}
\usepackage{amsmath}
\usepackage{amsfonts}
\usepackage{graphicx}
\usepackage{float}
\usepackage{setspace}
\usepackage{hyperref}
\usepackage{xkeyval}
\usepackage{multirow}
\usepackage{CJKutf8}

\begin{document}
\begin{CJK*}{UTF8}{bsmi}

\title{Optical phonons and magneto-elastic coupling in the ionic conductor AgCrSe$_2$}
\author{Jim Groefsema}
\affiliation{Institute of Physics, University of Amsterdam, Science park 904, 1098 XH Amsterdam, The Netherlands}
\author{Xuanbo Feng}
\affiliation{Institute of Physics, University of Amsterdam, Science park 904, 1098 XH Amsterdam, The Netherlands}
\affiliation{QuSoft, Science Park 123, 1098 XG Amsterdam, The Netherlands}
\author{Corentin Morice}
\affiliation{Institute of Physics, University of Amsterdam, Science park 904, 1098 XH Amsterdam, The Netherlands}
\author{Yingkai. Huang}
\affiliation{Institute of Physics, University of Amsterdam, Science park 904, 1098 XH Amsterdam, The Netherlands}
\author{Erik van Heumen}\email{e.vanheumen@uva.nl}
\affiliation{Institute of Physics, University of Amsterdam, Science park 904, 1098 XH Amsterdam, The Netherlands}
\affiliation{QuSoft, Science Park 123, 1098 XG Amsterdam, The Netherlands}

\begin{abstract} 
AgCrSe$_2$ is an example of a super-ionic conductor that has recently attracted attention for its low thermal conductivity. Here we investigate the optical properties of AgCrSe$_2$ in the ordered phase between 14 K and 374 K using reflectivity experiments. The far infrared optical response is dominated by three phonon modes, while six interband transitions are observed in the visible range. From our analysis we find that the phonon parameters display an interesting temperature dependence around the Néel temperature, pointing to a small magneto-elastic coupling. In addition, the lifetimes of the modes indicate that three-phonon processes dominate and the optical phonons decay into low energy acoustic modes involved in the super- ionic transition. Finally, we detect a small free charge carrier response through the analysis of Fabry-Perot interference fringes in our reflectivity data. 
\end{abstract}
\maketitle
\end{CJK*}

\section{Introduction}
With an increasing demand for batteries, solid-state ionic conductors have gained in interest due to their higher energy density, a potential for fast charging and increased safety over commonly used Li-ion batteries \cite{ohno_materials_2020}. Current in solid-state ionic conductors is carried by mobile ions through hopping between vacant sites in the material, which are typically slow processes. This has led to the study of a small group of materials that partially `melt' above the so-called super-ionic transition. Above this transition, the energy landscape of a sub-lattice of ions in the structure changes such that ions can rapidly diffuse through correlated motion \cite{he_origin_2017}. As it turns out, super-ionic conductors also have excellent thermoelectric response, owing to a very low thermal conductivity resulting from the partially molten sub-lattice of ions \cite{bailey_potential_2017}.

One example of such a super-ionic conductor with excellent thermoelectric figure of merit is AgCrSe$_2$, which consists of a sub-lattice of Ag ions sandwiched between CrSe$_2$ layers \cite{Gascoin_cmat_2011}. At a temperature of 475 K, AgCrSe$_2$ transitions into its super-ionic phase where Ag ions become disordered \cite{Engelsman_JSSC_1973, Boukamp_ionic_1983, van_der_lee_anharmonic_1989}. The phase transition into the disordered phase is second order \cite{Boukamp_ionic_1983}, but diffusion of Ag ions already seems to start well below the transition \cite{Li_natmat_2018}. The diffusion of Ag ions is promoted by lowering of the energy barrier for the occupation of a second interstitial site in the Ag lattice and has been proposed to result in the breakdown of certain phonon modes \cite{ding_anharmonic_2020}. However, direct imaging of the occupation of interstitial sites by Ag ions with transmission electron microscopy suggests that slow diffusion is a result of disorder \cite{Xie_JMCC_2019}. At low temperature, the Cr spins order in a non-collinear anti-ferromagnetic order with antiferromagnetic stacking along the c-axis \cite{Engelsman_JSSC_1973} and fluctuations of this order persisting up to 200 K \cite{damay_localised_2016}. Recently, short range correlations and the interplay between spin-orbit coupling and magnetic order have been studied at low temperatures\cite{Baenitz:PRB2021,Takahashi:PRM2022}.

The low thermal conductivity is most likely determined by the phonon spectrum and several momentum resolved studies have indeed observed changes in the phonon spectrum with temperature \cite{damay_localised_2016, Li_natmat_2018, ding_anharmonic_2020}. An early far-infrared optical study on sintered samples focussed on the high temperature transition and its impact on the phonon spectrum, reporting changes in the TO-LO mode splitting \cite{wakamura_JPCS_1996}. More recently, indications of the important role played by a 3 meV phonon mode and a possible magneto-elastic coupling between the low energy phonons and anti-ferromagnetic fluctuations have been brought to light \cite{damay_localised_2016}. This lead us to revisit AgCrSe$_2$ using high resolution optical spectroscopy reflectivity experiments on single crystalline samples, with a particular focus on the optical phonon modes and their temperature dependence. In this letter, we report reflectivity measurements between 14 K and 374 K, covering the low temperature anti-ferromagnetic state, but not the high temperature super-ionic transition. Three phonon modes are observed in the far infrared region and we find evidence for a weak magneto-elastic coupling through a detailed temperature dependence of the phonon mode parameters. We report the interband optical response and compare these to electronic structure calculations. Finally, we observe Fabry-Perot interference fringes in the reflectivity data. From a detailed analysis of these fringes, we are able to deduce a small free charge contribution to the optical response resulting in a DC resistivity of 0.5 $\Omega\,cm$. 

\section{Experimental Methods}
Single crystal AgCrSe$_2$ samples were produced using the chemical vapor transport growth method. The resulting AgCrSe$_2$ samples have the approximate dimensions of $3 \times 3$ mm with an average thickness between 100 and 200 $\mu$m. The crystal is supported by a copper holder, which has the shape of a tapered cone with the flat top surface cut to the shape of the AgCrSe$_2$ sample. In this way, light reflected from the copper holder will not reach the detector. The holder was polished to a smooth finish and cleaned using ultrasonic cleaning with a series of solutions in the order of citric acid, acetone and ethanol. Conducting silver epoxy was used to glue the sample to the holder and was subsequently baked at 125$^{\circ}$C for 20 minutes.

AgCrSe$_2$ does not cleave and is too soft to polish, therefore the as-grown surface was used for experiments. Fortunately, these surfaces naturally have a mirror smooth finish that is suitable for reflectivity measurements. The reflectivity data was obtained using a VERTEX 80v FTIR spectrometer over the photon energy range from $4$ meV to $3$ eV, using different light sources and detectors. The temperature dependence was measured using cooling and heating cycles between 14 K and 374 K at a rate of 1.66 K per minute. The heating and cooling stages have been repeated multiple times for the different photon energy ranges to improve the signal-to-noise ratio. In order to obtain a reference spectrum, gold or silver is evaporated on the sample surface. A new measurement using the same parameters and procedures as used for the sample measurement was then performed to obtain the reference spectrum. While it is possible to cleave aluminum and gold off the sample, cleaving silver turns out to be rather difficult. Due to this more than two different samples from the same batch were used. The reflectivity obtained in this way agreed very well in the ranges of overlap between different samples.

Electronic structure calculations of AgCrSe$_{2}$ are reported using density functional theory and an all-electron full-potential linearized augmented plane-wave basis set as implemented in Elk \cite{elk}. We employed the generalized gradient approximation in the shape of the Perdew, Burke, and Ernzerhof functional \cite{perdew_generalized_1996}, including spin-orbit coupling. We use a predefined high-quality set of parameters, and a Monkhorst-Pack grid of $16 \times 16 \times 16$ \textbf{k}-points in the Brillouin zone, which were checked for convergence. Calculations were done in the primitive unit cell, with ferromagnetic spin order, using an experimentally-measured crystal structure \cite{Engelsman_JSSC_1973}. Ferromagnetic ordering was chosen since the antiferromagnetic coupling cannot be observed in single-unit-cell calculations and the change between ferromagnetic and antiferromagnetic groundstates is low enough to be neglected. We find a magnetic moment of $3 \mu_B$ \cite{gautam:SSC2002}. Based on these results, we calculated the optical conductivity within the random phase approximation without local field effects, using a shifted grid of $20 \times 20 \times 20$ $\textbf{k}$-points and a smearing width of 50 meV.

\begin{figure}
\includegraphics[width=0.95\columnwidth]{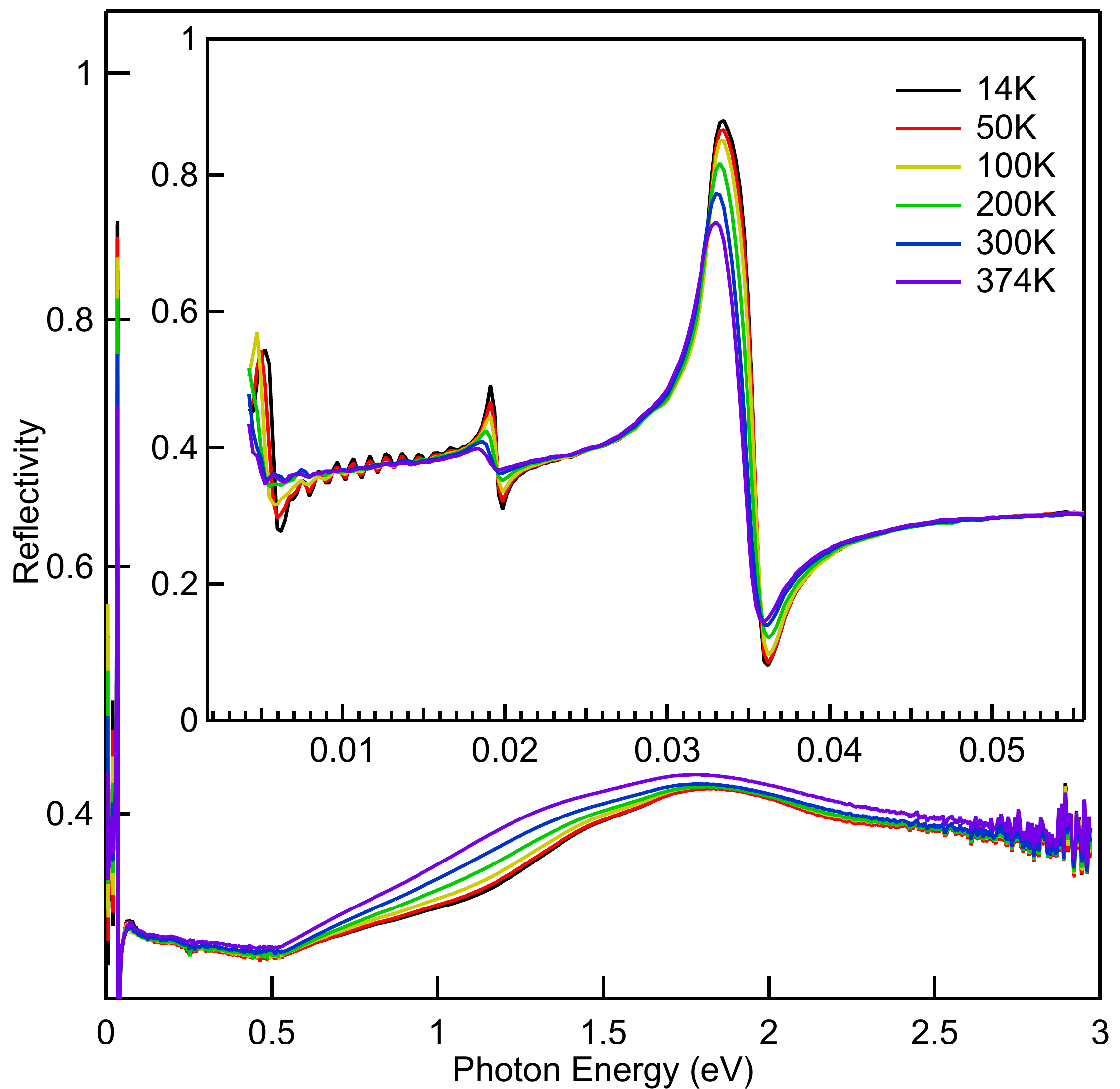}
\caption{The measured reflectivity against the photon energy used for selected temperatures. Interband transitions can be observed around $1$ eV and $1.5$ eV. The inset shows the reflectivity at a photon energy range of $4$ meV to $50$ meV. Three phonons are observed with smaller oscillations between $5$ and $15$ meV.}
\label{fig:FullR}
\end{figure} 

\section{Results}
Figure \ref{fig:FullR} shows the reflectivity against photon energy for a selection of temperatures. In the photon energy range above $0.5$ eV, the reflectivity shows several broad, temperature dependent structures that correspond to interband transitions. The inset of Fig. \ref{fig:FullR} shows the far-infrared (FIR) reflectivity data from $4$ meV to $50$ meV. Three structures are visible, corresponding to the three infrared active optical phonons expected for AgCrSe$_2$ \cite{wakamura_phonon_1990, wang_highly_2020}. Each of the modes shows significant temperature dependence, broadening as temperature is increased. As we will show in more detail below, these phonon modes can be well described by symmetric Lorentz modes, hinting at a weak electron-phonon coupling. The phonon modes appear to be unscreened by an electronic background (corresponding to free charge density) and the reflectivity spectrum resembles that of an insulator. This is further supported by a series of small oscillations that can be observed most clearly between $5$ and $15$ meV at low temperature. As we will show below these correspond to Fabry-Perot interference fringes and are a further indicator that  has very small free charge density. As temperature increases these fringes become weaker, but they remain visible up to the highest measured temperature.  

The first step in our analysis consists of creating a series of Drude-Lorentz models, one for each temperature, with parameters optimized using a least-squares optimization routine \cite{kuzmenko_kramerskronig_2005}. An accurate description of the reflectivity is obtained with a Drude-Lorentz model consisting of a total of nine Lorentz oscillators, describing three phonon modes and six interband transitions. The first interband transition appears in the mid infrared around 0.2 - 0.3 eV, while the other transitions are closer to the visible light range with an onset around 0.7 eV. The Drude-Lorentz model parameters for 14 K, 100 K, 300 K and 374 K are presented in table \ref{table:DLModel}.

\begin{figure}[h]
\includegraphics[width=0.95\columnwidth]{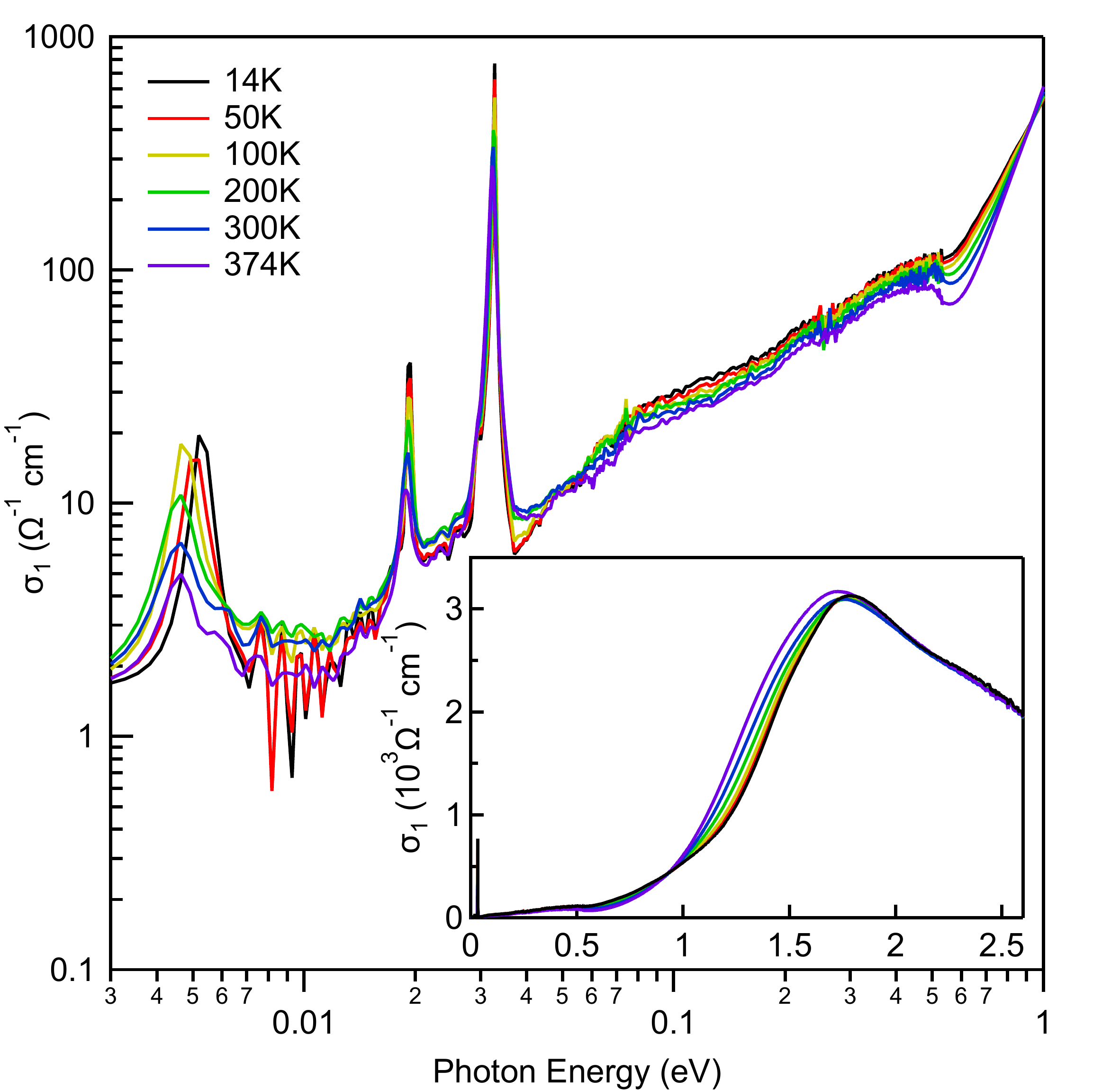}
\caption{The real part of the optical conductivity $\sigma_1 (\omega, T)$, plotted against photon energy on a log-log scale. We observe three phonon modes in the infrared (see table \ref{table:DLModel}), and at low temperature a number of Farby-Perot interference fringes (see main text for details). The inset shows the optical conductivity over the full energy range.}
\label{fig:S1}
\end{figure}
The Drude-Lorentz model provides the basis for the second step in our analysis. We used a variational dielectric function to effectively perform the Kramers-Kronig transformation and calculate the optical conductivity $\hat{\sigma} (\omega, T)$\cite{kuzmenko_kramerskronig_2005}. Figure \ref{fig:S1} shows $\sigma_1 (\omega, T)$ plotted against photon energy on a log-log scale from 3 meV up to 1 eV. The far infrared is characterized by three phonon modes on top of a small, but finite, background conductivity of a few $\Omega^{-1}\,cm$. The conductivity steadily starts to increase above 10 meV, possibly with some small structure around 0.1 eV. Above 0.5 eV, the conductivity starts to increase more rapidly. This energy gap is in good agreement with estimates based on static measurements \cite{Boukamp_ionic_1983}. The inset shows $\sigma_1 (\omega, T)$ on a linear photon energy scale up to 3 eV. 

\begin{figure}[h]
\includegraphics[width=0.95\columnwidth]{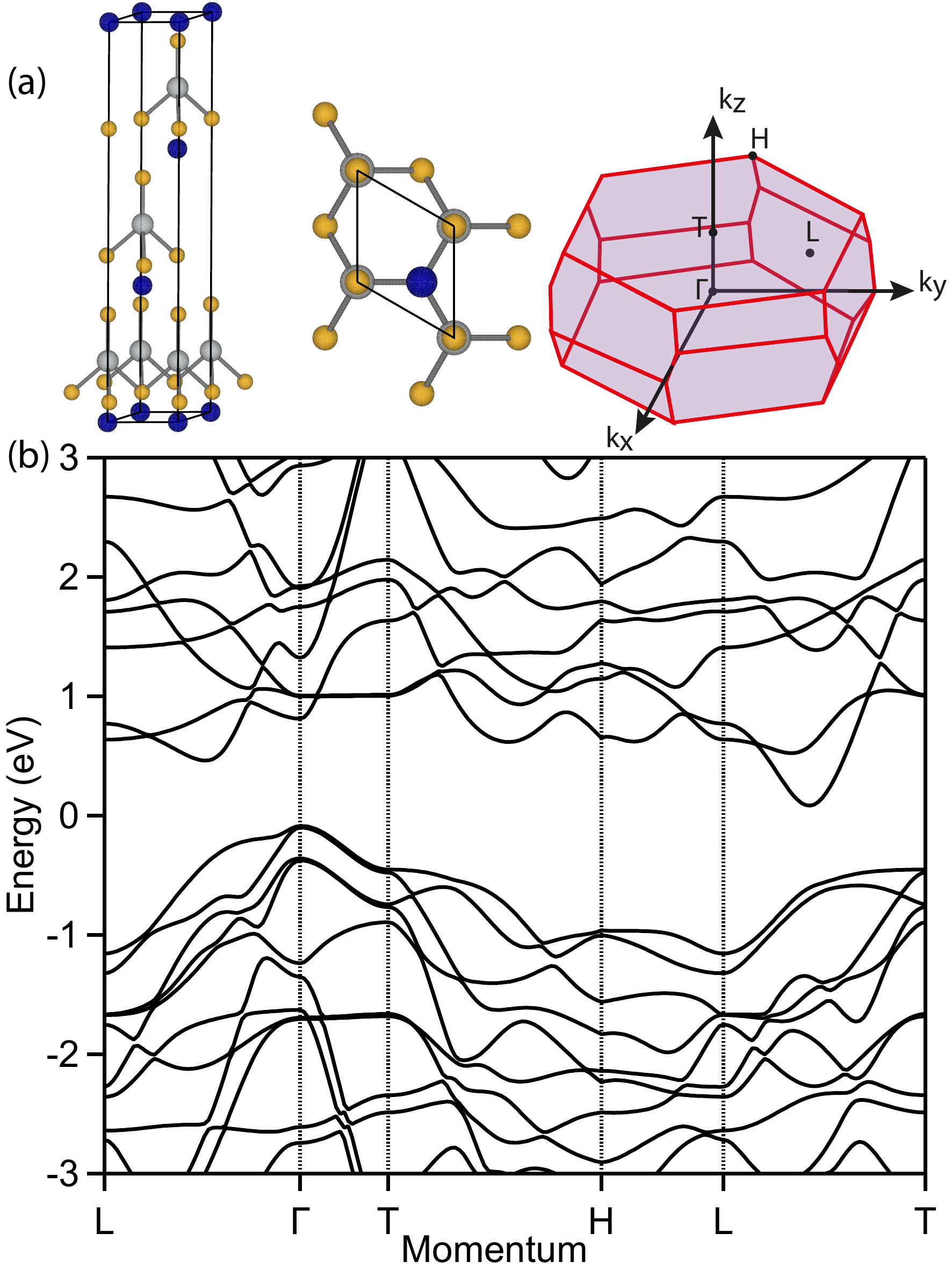}
\caption{(a): from left to right, (i) crystal structure viewed along the b-axis (Ag in blue, Cr in gray and Se in orange), (ii) crystal structure viewed along the c-axis and (iii) the Brillouin zone with several high symmetry points indicated. (b): the electronic bandstructure of AgCrSe$_2$ calculated along high symmetry lines. An indirect bandgap of 0.17 eV is observed, with the valence band maximum at $\Gamma$ and the conduction band minimum along the $L$-$T$ line.}
\label{fig:Bandstructure}
\end{figure}

\begin{figure}[t]
\includegraphics[width=0.98\columnwidth]{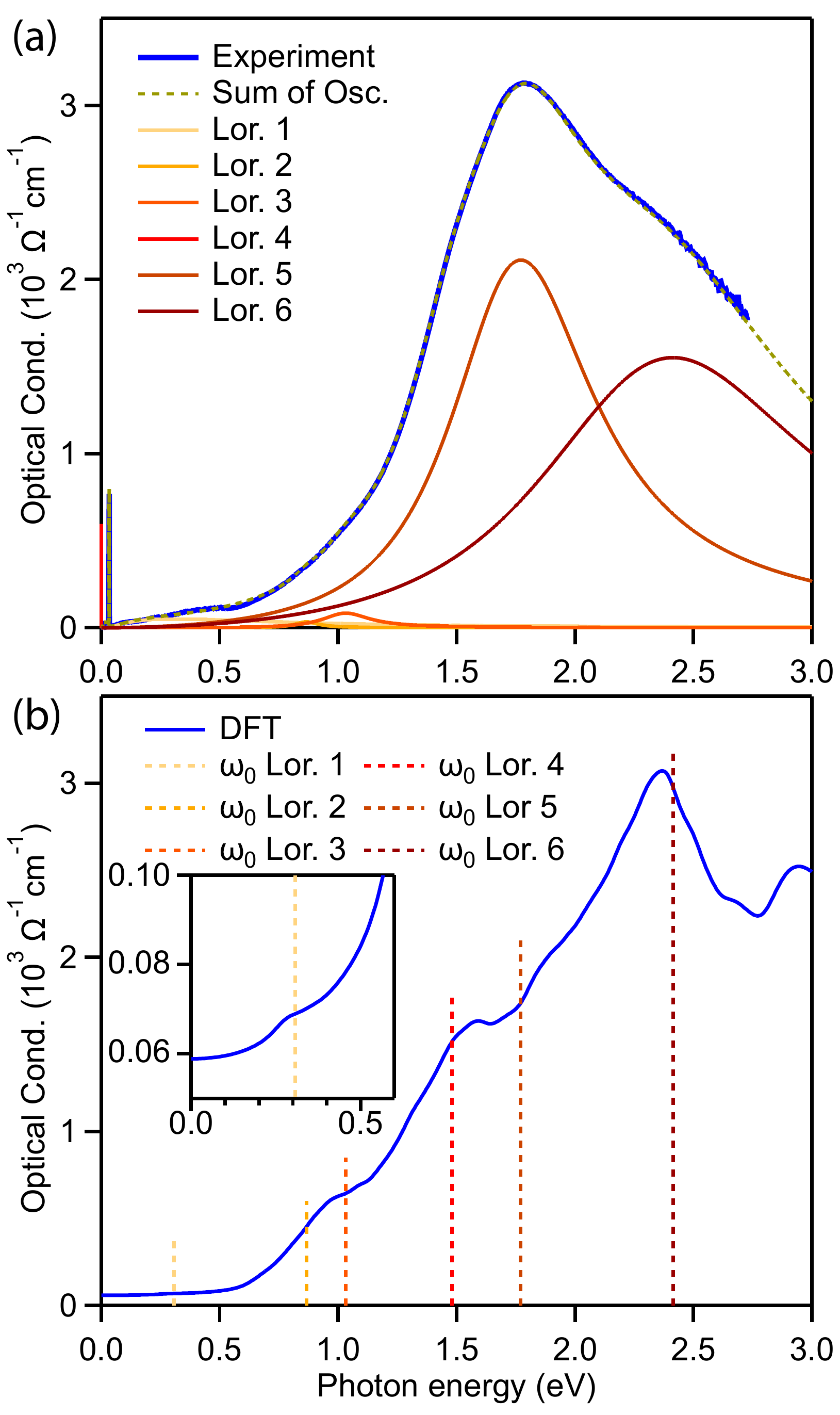}
\caption{(a): Experimental optical conductivity at 14 K and decomposition in interband transitions. Also shown is the sum of the individual oscillators (dashed yellow line) compared to the experimental data (blue line). (b): Comparison of the optical response calculated in the random phase approximation with experimental oscillator positions. Dashed lines indicate approximate interband transitions determined from the oscillators in (a). The inset shows that the transition associated with the indirect bandgap in the calculation around 0.2 eV corresponds well with the onset of interband transitions in the experimental data.}
\label{fig:compDFT}
\end{figure}

Figure \ref{fig:Bandstructure}a shows the unit cell and corresponding first Brillouin zone in which the high symmetry points are labelled. Fig. \ref{fig:Bandstructure} shows the electronic structure, calculated along several high symmetry lines of the Brillouin zone. The calculations point to a small indirect bandgap of about 0.17 eV. 
To compare the electronic structure calculations with our experiments, we compute the optical conductivity within the random phase approximation. The result is summarized in Fig. \ref{fig:compDFT}. Panel \ref{fig:compDFT}a shows the measured optical conductivity together with the Drude-Lorentz model fit. Also shown is the decomposition of the fit in individual oscillators that contribute to the conductivity. We use the resonance frequencies of these oscillators to compare to the calculated optical conductivity as presented in Fig. \ref{fig:compDFT}b. The calculated conductivity has slightly different overall shape, but qualitatively it seems to be in good agreement with the measured optical response. The energies of the experimentally obtained interband transitions (dashed vertical lines) closely correspond to transitions in the calculations. In particular, the lowest interband transition that is observed experimentally closely agrees with the onset of interband transitions in the calculation (see inset of Fig. \ref{fig:compDFT}b). This agreement suggests that the size of the indirect band gap obtained from the calculations is likely close to the actual value

\section{Discussion}
As mentioned in the introduction, the ionic conductivity in AgCrSe$_{2}$ is driven by delocalization of the Ag ions, which are sandwiched between CrSe$_{2}$ layers. At low temperature the Ag ions form a regular triangular lattice, while at high temperatures a second triangular lattice becomes partially occupied forming a honeycomb lattice. There are six infrared active phonons associated with this structure, of which three are in-plane modes \cite{wakamura_phonon_1990}. The phonon dispersions and partial phonon density of states of states have been previously calculated in Ref's \cite{Li_natmat_2018, Xie_PRL_2020}. These predict that IR active phonon modes may be expected around 5 meV, 17 meV and 30 meV \cite{Xie_PRL_2020}. The projected phonon density of states suggests that the lowest of these modes involves the motion of Ag ions, while the higher energy modes are mainly of Se character \cite{Li_natmat_2018}. 

\begin{figure}[h]
\includegraphics[width=0.98\columnwidth]{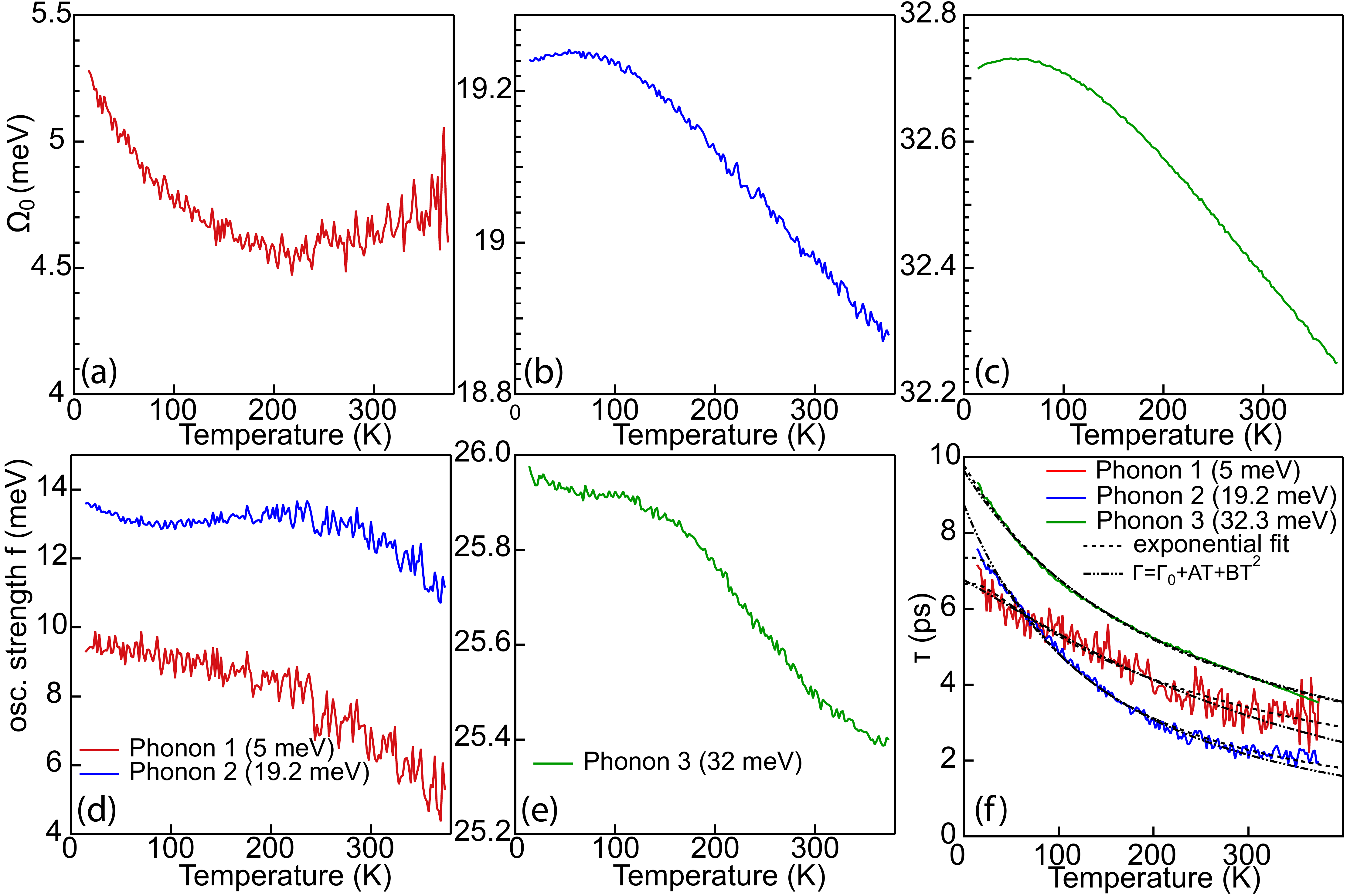}
\caption{Eigenfrequency $\Omega_0$ for phonon mode 1 (a), phonon 2 (b) and phonon 3 (c). Panels (d,e) display the corresponding oscillator strengths $f$. Panel (f) shows the lifetimes associated with the phonon modes as function of temperature together with fits to Eq. \ref{decay} (dashed black lines).}
\label{fig:PhononParam}
\end{figure}

Our Drude-Lorentz model provides direct access to the temperature dependent phonon parameters and they are displayed in Fig. \ref{fig:PhononParam}. The three phonon modes observed have eigenfrequencies of 5.2 (phonon 1), 19.2 (phonon 2) and 32.7 meV (phonon 3) at the lowest temperature, in good agreement with the calculations. The observation of the 19.2 meV mode corresponds well with the TO mode observed in Ref. [\citenum{Li_natmat_2018}]. In Fig. \ref{fig:PhononParam}a-c the temperature dependence of these modes is shown. As temperature increases, the modes soften as may be expected from increased thermal motion of the ions involved. 

The temperature dependent oscillator strengths of these modes are shown in Fig. \ref{fig:PhononParam}d,e. At high temperature the modes decrease in strength approximately as $f(T)=A-T^2$, which is typical for phonon modes. Interestingly, the phonon modes 2 and 3 show a small but marked increase in strength below approximately 60 K. This coincides with the temperature where the eigenfrequency reaches a maximum and is close to the Néel temperature at 55 K. Given that these modes have significant Se character and some Cr character, this could point to a small magneto-elastic coupling that was inferred from the correlation of negative thermal expansion with the onset of anti-ferromagnetic order \cite{damay_localised_2016}. The absence of an enhancement for phonon mode 1, which is of mostly Ag character, further supports the interaction of spin and lattice. 

Next, we turn to the phonon line widths, which at the lowest temperature are all close to being resolution limited and approximately 4 cm$^{-1}$ (or 8.3 ps). As temperature increases, these line widths increase and reach values at 374 K of 8 cm$^{-1}$  (4.2 ps), except for the 19.2 meV mode that broadens to 16 cm$^{-1}$ (2.1 ps). Previous works have focussed on the lifetime associated with the (acoustic) phonon modes in order to determine the origin of the low thermal conductivity \cite{Li_natmat_2018,Xie_JMCC_2019, Xie_PRL_2020}. In particular, in Ref \cite{Xie_PRL_2020} the authors focussed on the importance of 4-phonon scattering processes over 3-phonon scattering. From our detailed temperature dependent measurements, we can extract the lifetime of the IR active modes. Fig. \ref{fig:PhononParam}f shows the lifetime and compared to the 3-phonon plus 4-phonon scattering mechanism proposed in Ref. [\citenum{Xie_PRL_2020}] (dash-dotted lines). Based on these fits we find that 4-phonon scattering starts to dominate over 3-phonon scattering around 370 K for the lowest energy phonon mode (labelled with 5 meV). For the other phonon modes, we find that the 3-phonon scattering dominates at all temperatures below the order-disorder transition temperature. This suggests that for AgCrSe$_{2}$ the 3-phonon scattering dominates and we therefore focus on the more detailed calculations of the temperature dependence proposed in Ref's \cite{bairamov_1975, Anand_physB_1996}, 
\begin{equation}\label{decay}
\Gamma(T)=\Gamma_{0}+A\left(1+\frac{2}{e^{(E_{0}-E_{1})/k_{B}T}-1}+\frac{2}{e^{E_{1}/k_{B}T}-1}\right)
\end{equation}
where $\Gamma_{0}$ and $A$ are the intrinsic line width due to disorder and an anharmonic constant. The scattering process describes the decay of an optical phonon into two modes with final energies $E_{1}$ and $E_{0} - E_{1}$. 

We use Eq. \ref{decay} to fit the temperature dependent lifetimes of the three phonon modes in Fig. \ref{fig:PhononParam}. For this fit we use the intrinsic phonon energies $E_{0}$, observed in experiment (5.2 meV, 19.2 meV and 32.7 meV). We then obtain the phonon energy scales $E_{1}\approx$\, 3.0 meV, 12.7 meV and 32.5 meV for phonon 1, 2 and 3 respectively. The differences are approximately $E_{0} - E_{1}\approx$\,2 meV, 6.5 meV and 0.1 meV. Given that these energies are well below the optical branches, the above analysis indicates that the optical phonons decay into the low energy acoustic modes that are believed to be involved in the super-ionic transition \cite{Li_natmat_2018}. The comparison with the data is somewhat better at high temperatures, where the decrease in the lifetime slows down significantly faster than expected from 4-phonon scattering processes. 

To conclude our discussion, we briefly return to the small oscillations observed in the reflectivity data. Our initial analysis assumes that our crystal is optically infinitely thick. However, the free charge density is very small and according to previous work and our own bandstructure calculations, AgCrSe$_2$ should be insulating. If the screening is small and the sample thin, light can propagate through the sample and reflect of the copper sample holder supporting the sample. This reflected light can interfere with the light reflecting of the sample surface at the first reflection, giving rise to Fabry-Perot interference fringes. These fringes contain additional information on the optical properties of the sample as their amplitude and period are determined by the dielectric function. For this purpose, we use the original Drude-Lorentz model presented in Table \ref{table:DLModel} as input for the dielectric model of the AgCrSe$_2$ sample. For the Cu holder, we assume that it is infinitely thick and has a simple Drude response with typical parameters for a good metal: $\omega_{p}$\,=\,4 eV and $\Gamma$\,=\,0.07 eV. These two models are used as input for a multi-layer dielectric function model from which we obtain the reflectivity. 

\begin{figure}[h]
\includegraphics[width=0.8\columnwidth]{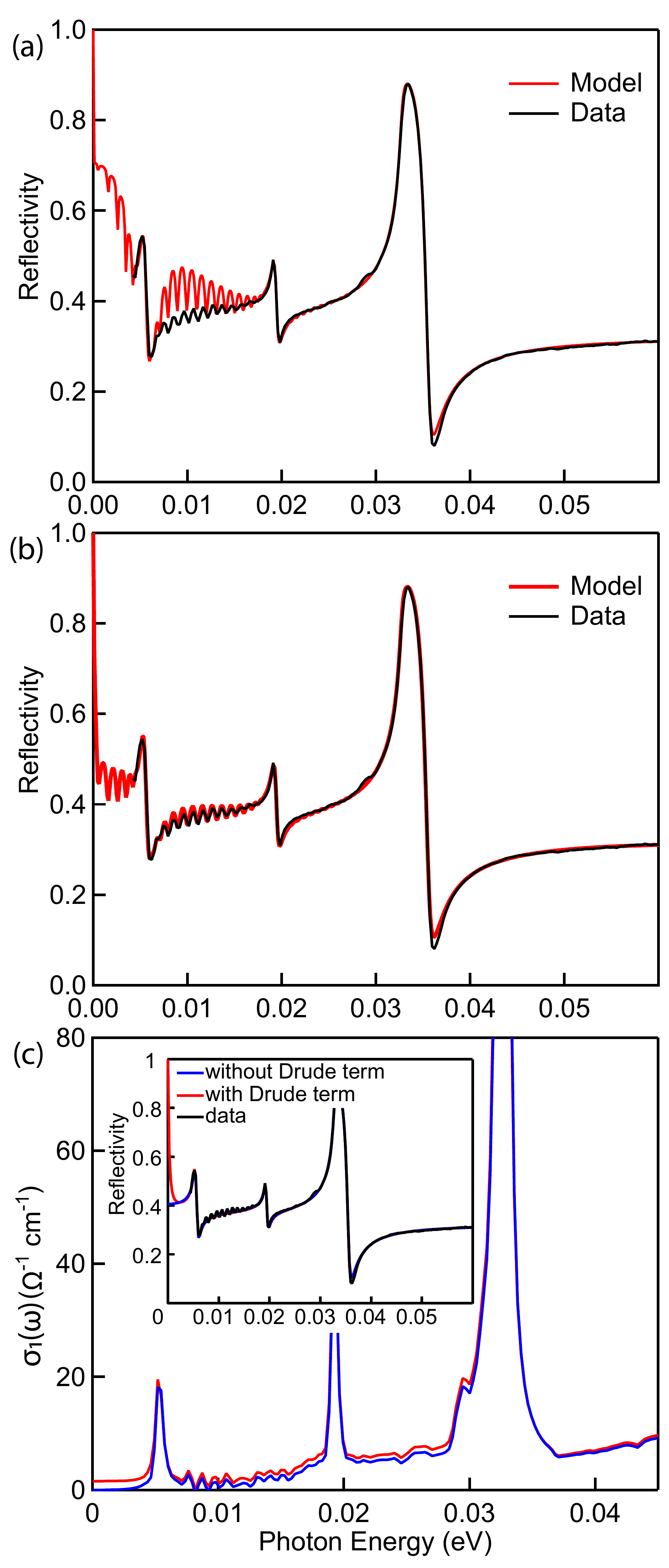}
\caption{(a): Model calculation of Fabry-Perot interference pattern compared to experimental data based on the DL model obtained from initial fits. (b): The same calculation, but now with an additional Drude response added to the model. (c): Optical conductivity with and without Drude response. The inset shows the difference between the two models imprinted on the reflectivity.}
\label{fig:DL-Model}
\end{figure}

Figure \ref{fig:DL-Model}a shows the comparison between the multi-layer model and the measured reflectivity at 14 K. To get the period of the model to agree with the data we used a sample thickness of 130 $\mu$m, which is consistent with measurements of the sample. Although the period of the oscillations matches well with the data, there is a significant discrepancy in the overall shape of the reflectivity. This is due to higher order reflections contributing and resulting in a beating pattern. We found that the only way to suppress these beatings was to introduce a small Drude response to the Drude - Lorentz model of the AgCrSe$_2$ crystal. The corresponding model is shown in figure \ref{fig:DL-Model}\textbf{c}, while the parameters of the Drude term are listed in Table \ref{table:DLModel}. This small Drude response introduces a finite DC conductivity of approximately 2 $\Omega^{-1} cm^{-1}$, which is smaller than found in previous studies \cite{Gascoin_cmat_2011}. The finite conductivity is likely linked to Ag vacancies resulting in a depletion of the valence band. The low conductivity therefore points to only a small Ag vacancy and good quality of our crystal. The plasma frequency obtained from our fit corresponds to a carrier density $n_{eff}\,\approx\,1.1\times\,10^{17}$\,cm$^{-3}$, placing it well below the regime where a large positive magnetoresistance was observed\cite{Takahashi:PRM2022}. Finally we note that the presence of the Drude response could not have been detected based on our reflectivity data alone. As the inset of \ref{fig:DL-Model}\textbf{c} shows, the original Drude-Lorentz model and the model with Drude response give the same level of agreement with our experimental data. 

\section{Summary}
To summarize our findings, we have measured the reflectivity for AgCrSe$_2$ over the photon energy range of $4$ meV to $3$ eV between 14 K and 375 K. Three phonon modes are observed with eigenfrequencies of 5.2 meV, 19.2 meV and 32.7 meV at low temperature. These modes soften slightly as temperature increases. The oscillator strength shows a small increase with decreasing temperature close to the Néel temperature, possibly indicating a weak magneto-elastic coupling. The temperature dependence of the line width suggests that 4-phonon scattering is less important then 3-phonon scattering in the temperature range of our experiments. Assuming that 3-phonon processes dominate, the temperature dependence is compatible with coupling to low energy acoustic modes. The optical gap is approximately 0.5 eV and we have reported three interband transition in the visible light part of the spectrum. Finally, we observe Fabry-Perot interference fringes in the reflectivity data that enable us to determine a free carrier density that provides a small contribution to the DC conductivity.

\section{acknowledgement}
The authors thank the Institute of Physics and H. Ellermeijer for continued support. This work is supported by the research center for quantum software, QuSoft.

\setlength{\tabcolsep}{22pt}
\begin{table*}[t]
\centering
\resizebox{0.85\textwidth}{!}{
\begin{tabular}{c|cccccc}
\hline
Temperature&&14\,K&100\,K&300\,K&374\,K\\
\hline
\multirow{2}{*}{Drude*}
&$\omega_p$&12.03&12.03&12.03&12.03\\
&$\gamma_D$&12.40&12.40&12.40&12.40\\
\hline
\multirow{3}{*}{Phonon 1}&$\Omega_{0}$&5.28&4.74&4.59&4.60\\
&$f$&9.28&9.35&7.10&5.28\\
&$\gamma$&0.58&0.71&1.33&1.13\\
\hline
\multirow{3}{*}{Phonon 2}&$\Omega_{0}$&19.24&19.23&18.98&18.88\\
&$f$&13.58&13.09&13.08&11.15\\
&$\gamma$&0.55&0.86&2.01&2.09\\
\hline
\multirow{3}{*}{Phonon 3}&$\Omega_{0}$&32.72&32.71&32.39&32.25\\
&$f$&51.95&51.85&51.01&50.80\\
&$\gamma$&0.44&0.61&0.98&1.17\\
\hline
\multirow{3}{*}{Lorentz 1}&$\Omega_{0}$&306&280&224&199\\
&$f$&574&518&394&320\\
&$\gamma$&879&850&643&459\\
\hline
\multirow{3}{*}{Lorentz 2}&$\Omega_{0}$&866&1284&1328&1335\\
&$f$&178&625&1054&1167\\
&$\gamma$&128&340&281&264\\
\hline

\multirow{3}{*}{Lorentz 3}&$\Omega_{0}$&1032&1022&1132&1152\\
&$f$&392&403&617&754\\
&$\gamma$&245&250&243&233\\
\hline
\multirow{3}{*}{Lorentz 4}&$\Omega_{0}$&1480&1487&1514&15245\\
&$f$&1452&1412&1341&1422\\
&$\gamma$&475&425&356&349\\
\hline
\multirow{3}{*}{Lorentz 5}&$\Omega_{0}$&1771&1771&1771&1771\\
&$f$&3422&3418&3370&3365\\
&$\gamma$&745&745&745&745\\
\hline
\multirow{3}{*}{Lorentz 6}&$\Omega_{0}$&2413&2413&2413&2413\\
&$f$&4066&4066&4066&4066 \\
&$\gamma$&1432&1432&1432&1432\\
\hline

\end{tabular}
}
\caption{Parameters for the optimized Drude-Lorentz model at selected temperatures. All parameters are in meV. $\varepsilon_\infty$ is set to $2.83$ for all temperatures. The Drude term makes use of the parameters $\omega_p$ and $\gamma_D$, corresponding to the plasma frequency and scattering rate respectively. The other parameters presented are the eigenfrequency $\Omega_0$, scattering rate $\gamma$ and the oscillator strength $f$. (*): The parameters of the Drude term are obtained from the analysis of the Fabry-Perot fringes of the 14 K data and have not been optimized at other temperatures.}
\label{table:DLModel}
\end{table*}

\bibliography{AgCrSe2_paper}
\end{document}